\documentclass[12pt]{iopart}

\usepackage{graphicx}

\newcommand{\DtwoD}{D}
\newcommand{\DctwoD}{D_{\rm c}}
\newcommand{\DthreeD}{D^{({\rm 3D})}}
\newcommand{\DcthreeD}{D_{\rm c}^{({\rm 3D})}}
\newcommand{\NctwoDid}{N_{\rm c}^{({\rm id})}}
\newcommand{\NcthreeDid}{N_{\rm c}^{({\rm 3D,id})}}
\newcommand{\Nclatid}{N_{\rm c}^{\rm (lat,id)}}
\newcommand{\NctwoDmf}{N_{\rm c}^{\rm (mf)}}
\newcommand{\NctwoDQMC}{N_{\rm c}^{\rm (QMC)}}
\newcommand{\Ncexp}{N_{\rm c}^{\rm (exp)}}
\newcommand{\ntwo}{n}
\newcommand{\rhotwo}{\rho}
\newcommand{\nthree}{n_3}
\newcommand{\gtwoD}{g}
\newcommand{\gthreeD}{g^{\rm (3D)}}
\newcommand{\TctwoDid}{T_{\rm c}^{\rm (id)}}
\newcommand{\Tcexp}{T_{\rm c}^{\rm (exp)}}
\newcommand{\aoh}{a_{\rm ho}}

\begin{document}

\title[The trapped two-dimensional Bose gas]{The trapped
two-dimensional Bose gas:\\ from Bose-Einstein condensation to
Berezinskii-Kosterlitz-Thouless physics}

\author{Z. Hadzibabic$^{1,2}$, P. Kr\"uger$^{1,3,4}$, M. Cheneau$^1$,
S. P. Rath$^1$ and J. Dalibard$^1$}
\date{}
\address{$^1$ Laboratoire Kastler Brossel, CNRS, \'Ecole normale sup\'erieure,
24 rue Lhomond, 75005 Paris, France}

\address{$^2$ Cavendish Laboratory, University of Cambridge, Cambridge CB3 0HE,
United Kingdom}

\address{$^3$ Kirchhoff Institut f\"ur Physik, Universit\"at Heidelberg, 69120
Heidelberg, Germany}

\address{$^4$ Midlands Centre for Ultracold Atoms, School of Physics and
Astronomy, University of Nottingham, Nottingham NG7 2RD, United
Kingdom.}

\ead{jean.dalibard@lkb.ens.fr}

\begin{abstract}
We analyze the results of a recent experiment with bosonic rubidium
atoms harmonically confined in a quasi-two-dimensional geometry. In
this experiment a well defined critical point was identified, which
separates the high-temperature normal state characterized by a
single component density distribution, and the low-temperature state
characterized by a bimodal density distribution and the emergence of
high-contrast interference between independent two-dimensional
clouds.  We first show that this transition cannot be explained in
terms of conventional Bose-Einstein condensation of the trapped
ideal Bose gas. Using the local density approximation, we then
combine the mean-field (MF) Hartree-Fock theory with the prediction
for the Berezinskii-Kosterlitz-Thouless transition in an infinite
uniform system. If the gas is treated as a strictly 2D system, the
MF predictions for the spatial density profiles significantly
deviate from those of a recent Quantum Monte-Carlo (QMC) analysis.
However when the residual thermal excitation of the strongly
confined degree of freedom is taken into account, an excellent
agreement is reached between the MF and the QMC approaches. For the
interaction strength corresponding to the experiment, we predict a
strong correction to the critical atom number with respect to the
ideal gas theory (factor $\sim 2$). A quantitative agreement between
theory and experiment is reached concerning the critical atom number
if the predicted density profiles are used for temperature
calibration.
\end{abstract}
\maketitle

\parindent 0mm

\section{Introduction}

As first noticed by Peierls \cite{Peierls:1935}, collective physical
phenomena in an environment with a reduced number of dimensions can
be dramatically changed with respect to our experience in three
dimensions. The example of Bose-Einstein condensation in a uniform
gas is a good illustration of the crucial role of dimensionality. In
three dimensions (3D), condensation occurs at a finite temperature,
and the phase of the macroscopic wave function exhibits long range
order \cite{Penrose:1956}. In two dimensions, such long range order
is destroyed by thermal fluctuations at any finite temperature, both
for an ideal and for an interacting Bose gas
\cite{Hohenberg:1967,Mermin:1966}.

In presence of repulsive interactions between particles, a uniform
2D Bose gas can nevertheless undergo a phase transition from a
normal to a superfluid state at a finite critical temperature. This
transition was predicted by Berezinskii \cite{Berezinskii:1971} and
by Kosterlitz and Thouless \cite{Kosterlitz:1973} (BKT), and it has
been observed in several macroscopic quantum systems, such as helium
films adsorbed on a substrate \cite{Bishop:1978}. The superfluid
state exhibits quasi-long range order, such that the one-body
correlation function decays algebraically at large distance. By
contrast the decay is exponential in the normal phase.

The recent advances in the manipulation of quantum atomic gases have
made it possible to address the properties of low-dimensional Bose
gases with novel tools and diagnostic techniques
\cite{Gorlitz:2001b,Rychtarik:2004,Smith:2005,Colombe:2004,Burger:2002,Kohl:2005b,Orzel:2001,Spielman:2007,Hadzibabic:2004,Hadzibabic:2006,Kruger:2007}
(for recent reviews, see \cite{Posazhennikova:2006,Bloch:2007}). A
recent cold atom experiment also addressed the BKT problem by
realizing a two-dimensional array of atomic Josephson junctions
\cite{Schweikhard:2007}. All these systems bring new questions,
since one is now dealing with a harmonically trapped, instead of a
uniform, fluid. In particular, due to a different density of states,
even in 2D one expects to recover the Bose-Einstein condensation
phenomenon in the ideal Bose gas case \cite{Bagnato:1991}. The total
number of atoms in the excited states of the trap is bounded from
above, and a macroscopic population of the ground state appears for
large enough atom numbers. However real atomic gases do interact. It
is therefore a challenging question to understand whether in
presence of atomic interactions, a trapped Bose gas will undergo a
BKT superfluid transition like in the uniform case, or whether
conventional Bose-Einstein condensation will take place, as for an
ideal system.

In recent experiments performed in our laboratory
\cite{Hadzibabic:2006,Kruger:2007}, a gas of rubidium atoms was
trapped using a combination of  a magnetic trap providing harmonic
confinement in the $xy$ plane, and an optical lattice, ensuring that
the third degree of freedom ($z$) of the gas was frozen. The
analysis of the atomic density profile revealed a critical point,
between a high temperature phase with a single component density
distribution, and a low temperature phase with a clear bimodal
distribution \cite{Kruger:2007}. This critical point also
corresponded to the onset of clearly visible interferences between
independent gases, which were used to study the coherence properties
of the system \cite{Hadzibabic:2006}. Surprisingly, the density
profile of the normal component was observed to be close to a
gaussian all the way down (in temperature) to the critical point.
This density profile is strikingly different from the one expected
for the ideal gas close to the BEC critical temperature.
Furthermore, if the width of the observed quasi-gaussian
distribution is interpreted as an empirical measure of the
temperature, this leads to a critical atom number at a given
temperature which is about five times larger than that needed for
conventional Bose-Einstein condensation in the ideal gas. These two
facts showed that, in sharp contrast to the 3D case, interactions in
2D cannot be treated as a minor correction to the ideal gas BEC
picture, but rather qualitatively change the behavior of the system.

The main goal of the present paper is to analyze this critical
point. We start in  \S~\ref{sec:ideal2D} with a brief review of the
properties of an ideal Bose gas in the uniform case and in the case
of harmonic confinement. In \S~\ref{sec:ideallattice} we adapt the
ideal gas treatment to the experimental geometry of
\cite{Kruger:2007}, and provide a detailed calculation showing that
the experimental results cannot be explained by this theory. Next,
in \S~\ref{sec:interactions}, we take interactions into account at
the mean-field (MF) level and we combine this analysis with the
numerically known threshold for the BKT transition in the uniform
case \cite{Prokofev:2001}. We first present a MF treatment for a
strictly 2D gas. For the parameter range explored experimentally, it
leads to a critical atom number in good agreement with the
prediction of the most recent Quantum Monte Carlo (QMC) calculations
\cite{Holzmann:2007b}. However the predicted MF density profiles
significantly differ from the QMC ones in the vicinity of the
critical point. In a second step, we take into account the residual
excitation of the $z$ motion in the mean-field model and we obtain
an excellent agreement with the QMC calculation. The predicted
density distribution near the critical point has a quasi-gaussian
shape and the ``empirical" temperature extracted from this
distribution is in fact somewhat lower than the real temperature.
Taking this into account we obtain good quantitative agreement
between experimental results and theoretical predictions. Finally we
summarize our findings and discuss the connection between the BEC
and the BKT transition in a 2D gas. While in a uniform, infinite
system only the latter can occur at finite temperature, in a trapped
gas both are possible, and the BEC transition can be viewed as a
special, non-interacting limit of the more general BKT behavior.

\section{Bose-Einstein condensation in an ideal 2D Bose gas}

\label{sec:ideal2D}

This section is devoted to a review of well known results concerning
the ideal Bose gas in two dimensions. We first address the case of a
uniform system at the thermodynamic limit, and we then consider a
gas confined in a harmonic potential.

\subsection{The uniform case}
\label{subsec:uniformideal}

In the thermodynamic limit a uniform, ideal Bose gas does not
undergo Bose-Einstein condensation when the temperature $T$ is
reduced, or the 2D spatial density $\ntwo$ is increased.
Bose-Einstein statistics leads to the following relation between the
phase space density $\DtwoD =\ntwo\lambda^2$ and the fugacity
$Z=\exp(\beta \mu)$
 \begin{equation}
\DtwoD =g_{1}(Z) \ ,\qquad g_\alpha(Z)=\sum_{j=1}^\infty
Z^j/j^{\alpha}\ .
 \label{eq:uniform2D}
 \end{equation}
Here $\lambda=\hbar(2\pi/(mk_BT))^{1/2}$ is the thermal wavelength,
$m$ is the atomic mass, $\beta=1/(k_BT)$ and $\mu$ is the chemical
potential. The function $g_\alpha(Z)$ is the polylogarithm, that
takes the simple form $ g_1(Z)=-\ln(1-Z)$ for $\alpha=1$. Because
$g_1(Z)\to +\infty$ when $Z\to 1$, (\ref{eq:uniform2D}) has a
solution in $Z$ for any value of $\DtwoD $. Hence no singularity
appears in the distribution of the population of the single particle
levels, even when the gas is strongly degenerate ($\DtwoD \gg 1$).
This is to be contrasted with the well known 3D case: the relation
$\DthreeD=g_{3/2}(Z)$ ceases to have a solution for a phase space
density above the critical value $\DcthreeD=g_{3/2}(1)\simeq 2.612$,
 where the 3D Bose-Einstein condensation phenomenon takes place.

\subsection{The ideal 2D Bose gas in a harmonic confinement}

\label{subsec:ideal2Dtrap}

We now consider an ideal gas confined in a harmonic potential
$V({\bf r})=m\omega^2r^2/2$. We assume that thermal equilibrium has
been reached, so that the population of each energy level is given
by Bose-Einstein statistics. Since the chemical potential $\mu$ is
always lower than the energy $\hbar \omega$ of the ground state of
the trap, the number of atoms $N'$ occupying the excited states of
the trap cannot exceed the critical value
 $\NctwoDid$
 \begin{equation}
\NctwoDid=\sum_{j=1}^{+\infty}\frac{j+1}{\exp(j\beta\hbar
\omega)-1}\ .
 \end{equation}
This expression can be evaluated in the  so-called
\emph{semi-classical limit} $k_BT\gg \hbar \omega$ by replacing the
discrete sum by an integral over the energy ranging from $0$ to
$+\infty$ \cite{Bagnato:1991}:
\begin{equation}
\NctwoDid=\left( \frac{k_BT}{\hbar\omega}\right)^2\;g_2(1)\ ,
 \label{eq:idealtrapped2}
 \end{equation}
with $g_2(1)=\pi^2/6$. This result also holds in the case of an
anisotropic harmonic potential in the $xy$ plane, in which case
$\omega$ is replaced by the geometric mean $\bar
\omega=\sqrt{\omega_x\omega_y}$, where $\omega_x,\omega_y$ are the
two eigenfrequencies of the trap. The saturation of the number of
atoms in the excited states is a direct manifestation of
Bose-Einstein condensation: any total atom number $N$ above
$\NctwoDid$ must lead to the accumulation of at least $N-
 \NctwoDid$ in the ground state of the trap.

Equation (\ref{eq:idealtrapped2}) is very reminiscent of the result
for the harmonically trapped 3D gas, where the saturation number is
$\NcthreeDid=(k_BT/(\hbar \omega))^3\;g_3(1)$. However an important
difference arises between the 2D and the 3D cases for the spatial
density profile. In 3D the phase space density in ${\bf r}$ is given
by $\DthreeD({\bf r})=g_{3/2}(Ze^{-\beta V({\bf r})})$ in the limit
$k_BT \gg \hbar \omega$. The threshold for Bose-Einstein
condensation is reached for $Z=1$; at this point $N$ is equal to the
critical number $\NcthreeDid$ and simultaneously the phase space
density at the center of the trap $\DthreeD(0)$ equals the critical
value $g_{3/2}(1)$. This allows for a simple connection between the
BEC thresholds for a homogenous gas and for a trapped system in the
semi-classical limit $k_BT \gg \hbar \omega$. In 2D such a simple
connection between global properties (critical atom number
$\NctwoDid$) and local properties (critical density at center
$\ntwo(0)$) does not exist. Indeed the semi-classical expression of
the  2D phase space density is
 \begin{equation}
\DtwoD ({\bf r})=g_1(Ze^{-\beta V({\bf r})})\ .
 \label{eq:2Didealphasespace}
 \end{equation}
Because $g_1(1)=+\infty$ this leads to a diverging value at the
center of the trap when $Z$ approaches 1. Therefore, although the
integral of $\DtwoD ({\bf r})$ over the whole space converges for
$Z=1$ and allows to recover (\ref{eq:idealtrapped2}), the
semiclassical result (\ref{eq:2Didealphasespace}) cannot be used to
derive a local criterion for condensation at the center of the trap.

One can go beyond the semi-classical approximation and calculate
numerically the central phase space density as a function of the
total number of atoms. We consider as an example the trap parameters
used in \cite{Kruger:2007}, where $\omega_x/(2\pi)=9.4$~Hz,
$\omega_y/(2\pi)=125$~Hz. In the typical case $k_BT/(\hbar \bar
\omega)=50$ ($T\simeq 80$~nK), the discrete summation of the
Bose-Einstein occupation factors for $Z=1$ gives $N_{\rm c} \simeq
4800$ (the value obtained from the semi-classical result
(\ref{eq:idealtrapped2}) is 4100). Using the expression of the
energy eigenstates (Hermite functions), we also calculate the phase
space density at the origin and we find $\DtwoD (0)\simeq 13$. Let
us emphasize that this value is a mere result of the finite size of
the system, and does not have any character of universality.

\section{Condensation of an ideal Bose gas in a harmonic + periodic potential}

\label{sec:ideallattice}

In order to produce a quasi two-dimensional gas experimentally, one
needs to freeze the motion along one direction of space, say $z$. In
practice this is conveniently done using the potential $V^{\rm
(lat)}(z)=V_0\;\sin^2(kz)$ created by an optical lattice along this
direction. A precise comparison between the measured critical atom
number and the prediction for an ideal gas requires to properly
model the confining potential and find its energy levels. This is
the purpose of the present section.

\subsection{The confining potential}

The optical lattice is formed by two running laser waves of
wavelength $\lambda_L$, propagating in the $yz$ plane with an angle
$\pm\theta/2$ with respect to the $y$ axis. The period
$\ell=\pi/k=\lambda_L/(2\sin(\theta/2))$ of the lattice along the
$z$ direction can be adjusted to any value above $\lambda_L/2$ by a
proper choice of the angle $\theta$. For a blue-detuned lattice
($\lambda_L$ is smaller than the atomic resonance wavelength), $V_0$
is positive and the atoms accumulate in the vicinity of the nodal
planes $z=0$, $z=\pm \pi/k$, etc. The oscillation frequency at the
bottom of the lattice wells is $\omega^{(\rm
lat)}_z=2\sqrt{V_0E_r}/\hbar$, where $E_r=\hbar^2k^2/(2m)$. In order
for the quasi-2D regime to be reached, $\hbar \omega^{(\rm lat)}_z$
must notably exceed the typical thermal energy $k_B T$ as well as
the interaction energy per particle for a non ideal gas.

For a blue detuned lattice an additional confinement in the $xy$
plane must be added to the optical lattice potential. This is
conveniently achieved using a magnetic trap, that creates a harmonic
potential with frequencies $\omega_x,\omega_y$. The magnetic trap
also provides an additional trapping potential $m\omega_z^2 z^2/2$
along the $z$ direction. The oscillation frequency $\omega_z$
created by the magnetic trap is usually much lower than the one
created by the lattice $\omega^{(\rm lat)}_z$. The main effect of
the magnetic confinement along the $z$ direction is to localize the
atoms in the ${\cal N}$ central lattice planes, where the effective
number of planes ${\cal N}\sim 4\;k_B T/(m\omega_z^2 \ell^2)$. As we
see below this number is on the order of 2 to 4 for the range of
parameters explored in \cite{Kruger:2007}. The fact that more than
just one plane is populated is an important ingredient of the
experimental procedure used in \cite{Hadzibabic:2006,Kruger:2007}.
It allows one to look for the interferences between planes, and to
access in this way the spatial coherence of the quasi-2D gas.

In order to extract thermodynamic information from the interference
between planes, one must ensure that the various populated planes
have the same temperature. This is achieved by using finite size
lattice beams in the $xy$ plane, so that atoms in the high energy
tail of the thermal distribution can actually travel quasi freely
from one plane to the other, thus ensuring thermalization. In the
experiments described in \cite{Hadzibabic:2006,Kruger:2007}, the
waist $W_x$ of the lattice beams along the $x$ direction was chosen
accordingly. The total trapping potential can then be written in the
following way
 \begin{equation}
 V({\bf r}) = V^{\rm (mag)}({\bf r})+V^{\rm (lat)}({\bf r})
 \label{eq:totalpot}
  \end{equation}
with
 \begin{eqnarray}
V^{\rm (mag)}({\bf r}) &=& \frac{1}{2}m\left( \omega_x^2 x^2+
\omega_y^2 y^2+\omega_z^2 z^2\right)\\
V^{\rm (lat)}({\bf r})&=&V_0\;e^{-2x^2/W_x^2}\;\sin^2(k(z-z_0))
 \label{eq:trapping_pot}
 \end{eqnarray}
Note that we have included here the offset $z_0$ between the optical
lattice and the bottom of the magnetic potential; this quantity was
not set to a fixed value in the experiments
\cite{Hadzibabic:2006,Kruger:2007}. We consider below two limiting
situations: (A) $kz_0=\pi/2$, with two principal equivalent minima
in $kz=\pm \pi/2$, (B) $kz_0=0$, with one principal minimum in $z=0$
and two side minima in $kz=\pm \pi$. At very low temperatures, we
expect that A will lead to two equally populated planes whereas
configuration B will lead to one populated plane.  For the
temperature range considered in practice, the differences between
the predictions for A and B are minor, as we will see below.

\subsection{Renormalization of the trapping frequency $\omega_x$ by the optical lattice}

\label{subsec:renorm}

In order to use the Bose-Einstein statistics for an ideal gas, one
needs to know the position of the single particle energy levels. For
the potential (\ref{eq:totalpot}) it is not possible to find an
exact analytical expression of these levels. However if the
extension of the atomic motion along the $x$ direction is smaller
than the laser waist, an approximate expression can be readily
obtained, as we show now.

The frequencies of the magnetic trap used in
\cite{Hadzibabic:2006,Kruger:2007} are $\omega_x=2\pi \times
10.6$~Hz and $\omega_y=\omega_z=2\pi\times 125$~Hz. The optical
lattice has a period $\pi/k=3\;\mu$m ($E_r=\hbar^2 k^2/(2m)=h\times
80$~Hz) and a potential height at center $U_0/h=35$~kHz
($1.7\;\mu$K). The lattice oscillation frequency at center ($x=0$)
is thus $\omega_z^{\rm (lat)}(x=0)=2\pi\times 3$~kHz
($\hbar\omega_z^{\rm (lat)}/k_B= 150$~nK).  When the atoms occupy
the ground state of the $z$ motion, they acquire the zero-point
energy $\hbar \omega_z^{\rm (lat)}(x)/2$ from the $z-$degree of
freedom. The dependance on $x$ of $\omega_z^{\rm (lat)}(x)$, due to
the gaussian term $e^{-2x^2/W_x^2}$ in the laser intensity, causes a
renormalization of the $x$ frequency:
 \begin{equation}
\omega_x^2 \rightarrow
{\omega'}_x^2=\omega_x^2-\frac{2\sqrt{V_0E_r}}{mW_x^2}\ .
 \label{eq:renormx}
 \end{equation}
The waist of the lattice beams is $W_x=120\,\mu$m which leads to
$\omega'_x=2\pi\times 9.4$~Hz. A similar effect should in principle
be taken into account for the frequency $\omega_y$. However the
scale of variation of the laser intensity along the $y$ axis is the
Rayleigh length, which is much larger than the waist $W_x$, and the
effect is negligible.

This simple way of accounting for the finiteness of the waist $W_x$
is valid when the extension of the motion along $x$ is small
compared to $W_x$. For $T=100$~nK, the width of the thermal
distribution along $x$ is $\sqrt{k_BT/m\omega_x^2}\sim 50\;\mu$m,
which is indeed notably smaller than $W_x$. Taking into account the
finiteness of $W_x$ by a mere reduction of the trapping frequency
along $x$ is therefore valid for the major part of the energy
distribution.

We note however that atoms in the high energy tail of the
distribution ($E>5\;k_BT$  for our largest temperatures) can explore
the region $|x|>W_x$, where the influence of the lattice beams is
strongly reduced.  In this region the atoms can move from one
lattice plane to the other. As explained above these atoms play an
important role by ensuring full thermalization between the various
planes. We now turn to an accurate treatment of the critical atom
number required for Bose-Einstein condensation, taking into account
these high energy levels for which the 2D approximation is not
valid.

\subsection{The critical atom number in a `Born--Oppenheimer' type approximation}

\label{subsec:Born}

In order to get the single particle energy eigenstates in the
lattice + harmonic potential confinement, and thus the critical atom
number, one could perform a numerical diagonalization of the 3D
hamiltonian with the potential (\ref{eq:totalpot}). This is however
a computationally involved task and it is preferable to take
advantage of the well separated energy scales in the problem.

\begin{figure}[ht]
\centerline{\includegraphics[width=8cm]{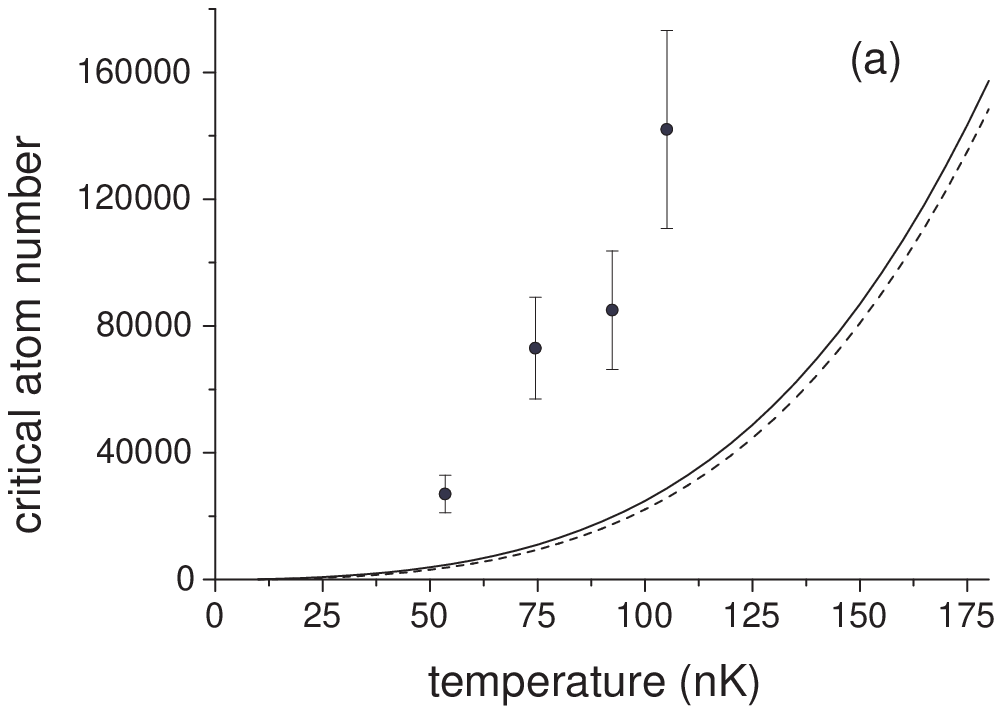}
\includegraphics[width=7cm]{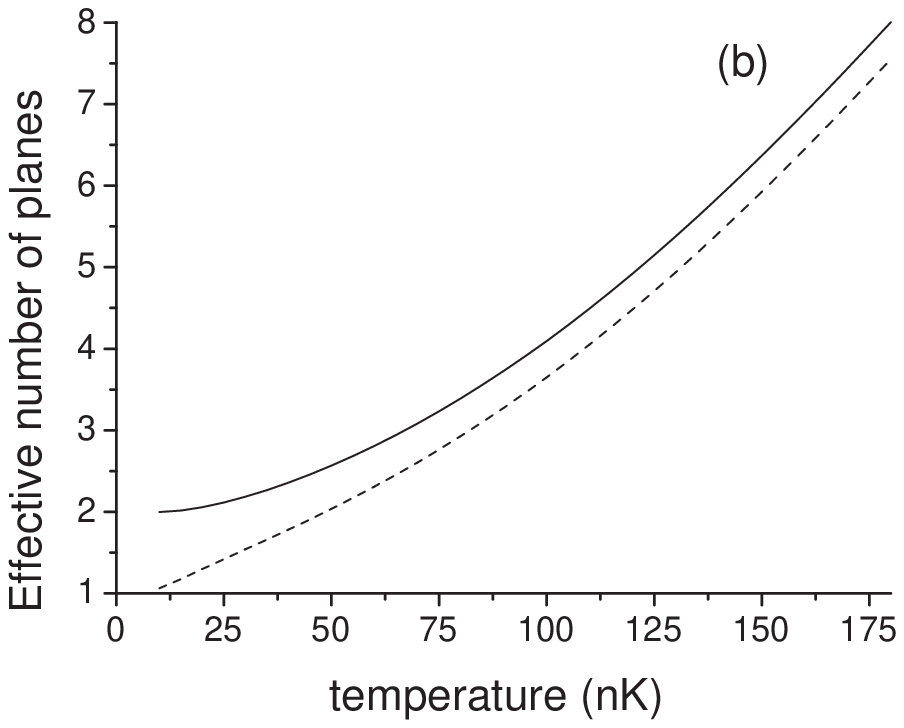}}
\caption{(a) Critical atom number $\Nclatid $ in the ideal gas model
for the optical lattice + magnetic trap configuration, as a function
of temperature. The points represent the experimental results of
\cite{Kruger:2007}, and the error bars combine the systematic and
statistical uncertainties on atom numbers. (b) Effective number of
planes ${\cal N}_{\rm eff}=\Nclatid /\NctwoDid$ as a function of
temperature. In both panels the continuous (dashed) line is for
configuration A (B). The calculation is performed using the first
100 eigenvalues of the $z$ motion, and the first neglected levels
$E_j(x)$ are 22~kHz ($1\;\mu$K) above the bottom of the trap.}
 \label{fig:Nc_lattice}
 \end{figure}

We first note that the trapping potential (\ref{eq:totalpot}) is the
sum of a term involving the variables $x$ and $z$, and a quadratic
component in $y$. The motion along the $y$ axis can then be
separated from the $xz$ problem, and it is easily taken into account
thanks to its harmonic character. For treating the $xz$ problem we
use a `Born-Oppenheimer' type approximation. We exploit the fact
that the characteristic frequencies of the $z$ motion are in any
point $x$ notably larger than the frequency of the $x$ motion. This
is of course true inside the lattice laser waist, since
$\omega^{(\rm lat)}_z /\omega_x \sim 300$, and it is also true
outside the laser waist as the $x$ direction corresponds to the weak
axis of our magnetic trap. Therefore we can proceed in two steps:
 \begin{enumerate}
 \item
For any fixed $x$ we numerically find the eigenvalues $E_j(x)$,
$j=0,1,\ldots$ of the $z$ motion in the $(V^{\rm (mag)}+V^{\rm
(lat)})(x,z)$ potential. We determine the $E_j$'s up to the
threshold $6\,k_BT$ above which the thermal excitation of the levels
is negligible. The result of this diagonalization  is shown in
figure \ref{fig:eigenvalues} for configurations $A$ and $B$.

 \begin{figure}[t]
\centerline{\includegraphics[width=7cm]{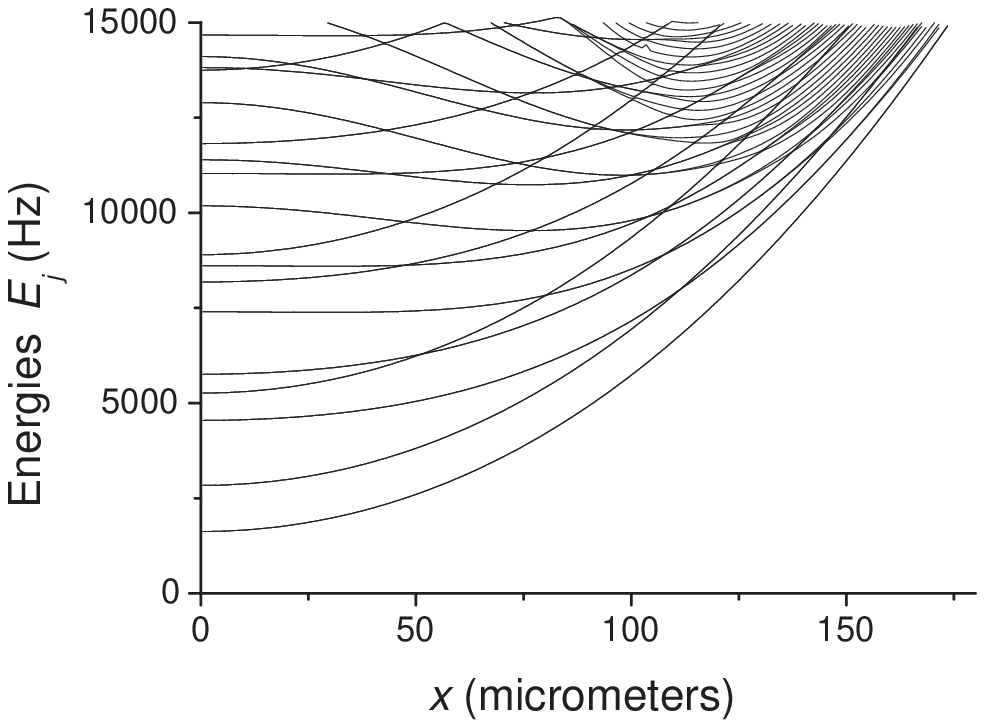}
\includegraphics[width=7cm]{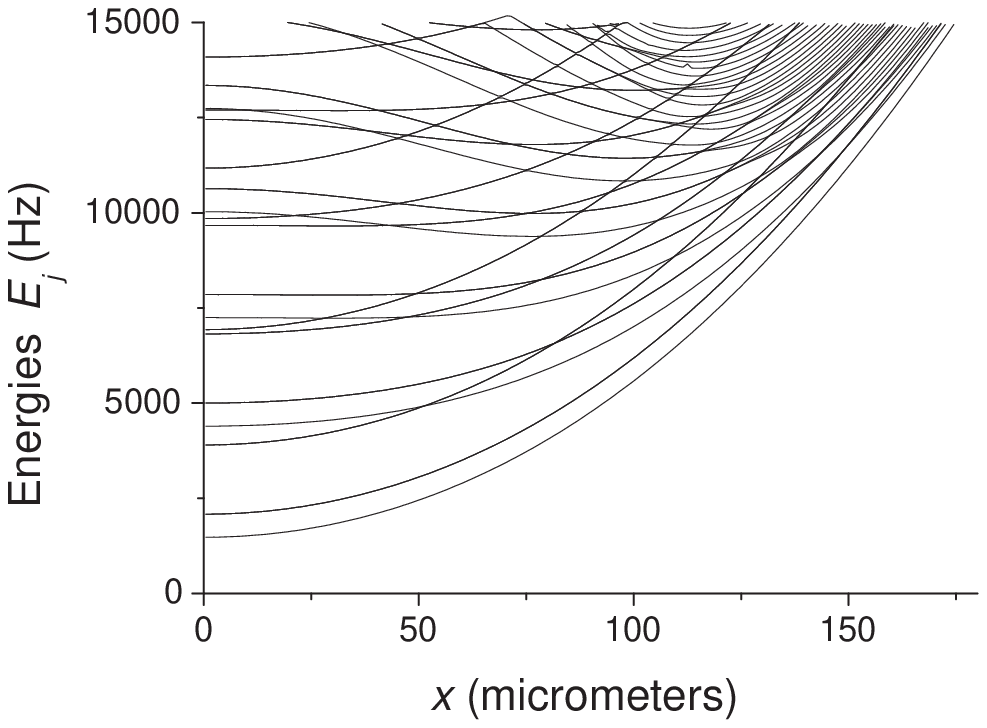}}
\caption{Eigenvalues of the $z$-motion in the magnetic+optical $x-z$
potential, for fixed values of the $x$ coordinate (left:
configuration A, right: configuration B).}
 \label{fig:eigenvalues}
 \end{figure}

 \item
We then treat semi-classically the $x$-motion on the various
potential curves $E_j(x)$. Adding the result for the independent
harmonic $y$ motion --also treated semi-classically-- we obtain the
surface density $\ntwo(x,y)$ (integral of the spatial density
$\nthree ({\bf r})$ along the direction $z$)
 \begin{equation}
\ntwo(x,y)=\frac{1}{\lambda^2}\sum_j\; g_1\left(
Ze^{-\beta(E_j(x)+V^{\rm (mag)}(y))} \right)\ .
 \label{eq:maglatresult}
 \end{equation}
 \end{enumerate}
This procedure yields a result that is identical to the
semi-classical prediction in two limiting cases:
 \begin{enumerate}

\item
The pure 2D case, that is recovered for large waists and low
temperatures. In this case the restriction to the closest-to-center
lattice plane and to the first $z$-level is legitimate, and the sum
over $j$ contains only one significant term corresponding to
(\ref{eq:2Didealphasespace}).

\item
the pure 3D harmonic case with zero lattice intensity where
$E_j(x)=m\omega_x^2 x^2/2+(j+1/2)\hbar\omega_z$. In this case the
sum over $j$ in (\ref{eq:maglatresult}) leads to $
\ntwo(x,y)\;\lambda^2=g_2\left(Ze^{-\beta V^{\rm
(mag)}(x,y)}\right)/(\beta \hbar \omega_z)$, which coincides with
the 3D result $\DthreeD({\bf r})=g_{3/2}(Ze^{-\beta V({\bf r})})$
when integrated along $z$.

\end{enumerate}
Of course this procedure also allows to interpolate between these
two limiting cases, which is the desired outcome. The integral of
$\ntwo$ in the $xy$ plane for $\mu={\rm min}\left(E_j(x)+V^{\rm
(mag)}(y)\right)$ gives the critical atom number $\Nclatid(T)$ in
the ideal gas model for this  lattice geometry. It is shown in
figure \ref{fig:Nc_lattice}a for the two configurations A and B.

The critical atom number $\Nclatid $ can be compared with the result
for a single plane $\NctwoDid$ with eigenfrequencies $\omega'_x$ and
$\omega_y$. The ratio gives the effective number of planes ${\cal
N}_{\rm eff}$, shown as a function of temperature in figure
\ref{fig:Nc_lattice}b for the two configurations A and B. This ratio
increases with temperature, which means that $\Nclatid$ increases
faster than $T^2$ with temperature in the temperature domain
considered here. For example in the range $50-110$~nK, the variation
of $\Nclatid $ is well represented by $T^\beta$, with $\beta=2.8$.

Three phenomena contribute significantly to this ``faster than
$T^2$" increase of $\Nclatid $. First in the lattice + harmonic
potential geometry, the number of contributing planes increases with
temperature, even if the atomic motion in each plane remains
two-dimensional (i.e.  the atom number per plane increasing strictly
as $T^2$). Second, we are exploring here a region of temperature
where $k_BT$ becomes non negligible with respect to $\hbar
\omega_z^{\rm (lat)}$ (the two quantities are equal for $T=150$~nK),
and the thermal excitations of the $z$-motion in each lattice plane
cannot be fully neglected. Third, for the largest considered
temperatures, the extension of the atomic motion along $x$ becomes
comparable to the laser waist, and the lattice strength is then
significantly reduced.

\subsection{Comparison with experimental results}
\label{subsec:comparison_exp}

In \cite{Kruger:2007} the critical atom number in the lattice +
magnetic trap configuration was measured for various ``effective"
temperatures, deduced from the width of the quasi-gaussian atomic
distribution. Each critical point ($\Ncexp,\Tcexp$) was defined as
the place where a bimodal spatial distribution appeared, if the atom
number was increased beyond this point at constant temperature, or
the temperature reduced at constant atom number. The critical point
also corresponded to the threshold for the appearance of
interferences with a significant contrast between adjacent planes.
The experimental measurements of critical points, taken over the
effective temperature range 50--110~nK, are shown as dots in figure
\ref{fig:Nc_lattice}a. The systematic + statistical uncertainties of
the atom number calibration are 25\%. Assuming that the effective
temperatures coincide with the true ones (this point will be
examined in \S~\ref{subsec:latticeMF}) we find $\Ncexp/\Nclatid \sim
5.3\;(\pm 1.2)$.

In addition to this large discrepancy between experiment and ideal
gas model for the critical atom numbers, one also finds a strong
mismatch concerning the functional shape of the column density $\int
\ntwo(x,y) dy$ that was measured in absorption imaging in
\cite{Hadzibabic:2006,Kruger:2007}. While the experimental result is
quasi-gaussian, the column density profiles calculated for an ideal
gas at the critical point are much `peakier'. An example is given in
the appendix for a single plane, and we checked that a similar shape
remains valid for our harmonic + lattice potential. We therefore
conclude that the experimental results of \cite{Kruger:2007} cannot
be accounted for with this ideal gas prediction for `conventional'
BEC.

\section{Interactions in a quasi 2D trapped Bose gas}

\label{sec:interactions}

To improve the agreement between the experimental results and the
theoretical modeling we now take repulsive atomic interactions into
account. In a first stage we present a 2D mean-field analysis, in
which the motion along $z$ is assumed to be completely frozen
whereas the $xy$ motion is treated semi-classically. In order to
model interactions in this case, we start from the 3D interaction
energy $(\gthreeD /2)\; \int \nthree ^2({\bf r})\;d^3r $, where
$\gthreeD =4\pi\hbar^2a/m$ and $a$ is the scattering length. The
$z$-degree of freedom is restricted to the gaussian ground state of
the confining potential, with an extension
$a_z=\sqrt{\hbar/(m\omega^{\rm (lat)}_z)}$, and the interaction
energy is
 \[
E_{\rm int}=\frac{\gtwoD}{2}\int \ntwo^2({\bf r})\;d^2r\ .
 \]
We set $\gtwoD=\hbar^2 \tilde g/m$, where the dimensionless
parameter $\tilde g=\sqrt{8\pi}\,a/a_z$ characterizes the strength
of the 2D interaction (for a more elaborate treatment of atomic
interactions in a quasi-2D geometry, see
\cite{Petrov:2000a,Petrov:2001}). For the optical lattice used in
\cite{Kruger:2007} we find $\tilde g=0.13$. In a second stage we
take into account the residual excitation of the $z$-motion in a
`hybrid' 3D mean-field approximation. We calculate in a
self-consistent way the quantum levels of the $z$ motion, whereas
the motion in the $xy$ plane is still treated semi-classically.

\subsection{Criticality within mean-field solutions: 3D vs. 2D}

We start our discussion with a brief reminder of the role of (weak)
interactions in a trapped 3D Bose gas \cite{Dalfovo:1999}. One often
uses the mean-field  Hartree-Fock approximation, that gives in
particular a relatively accurate value for the shift of the critical
temperature for Bose-Einstein condensation. In order to calculate
this shift, one assumes that above the critical temperature, the
atoms evolve in the effective potential $V_{\rm eff}({\bf r})=V({\bf
r})+2\gthreeD \nthree ({\bf r})$. The phase space density in ${\bf
r}$ is thus a solution of $\DthreeD({\bf r})=g_{3/2}(Ze^{-\beta
V_{\rm eff}({\bf r})})$. As for the ideal case this equation ceases
to have a solution when the central phase space density goes above
$g_{3/2}(1)$. The mere effect of repulsive interactions within the
mean-field approximation is to increase the number of atoms for
which this threshold is met. The increase is typically $\sim 10\%$
for standard trap and interaction parameters \cite{Dalfovo:1999}.

For a trapped 2D gas this treatment based on a local criterion
(phase space density at center) cannot be used. Indeed as explained
in section \ref{subsec:ideal2Dtrap}, it is not possible to identify
a critical phase space density at which BEC of the 2D gas is
expected. On the contrary the semiclassical approximation leads to
an infinite central density at the critical point, and it is unclear
whether one can achieve an arbitrarily large spatial density in
presence of repulsive interactions.

One could also look for a global criterion for criticality  based on
the total atom number. The starting point is the solution of the
mean-field equation
 \begin{equation}
\DtwoD({\bf r}) = g_1\left(Ze^{-\beta V_{\rm eff}({\bf r})} \right)
 \label{eq:2DHartreeFock}
 \end{equation}
with $V_{\rm eff}({\bf r})=V({\bf r})+2\gtwoD\, \ntwo({\bf r})$.
When $\gtwoD =0$, we saw in \S \ref{sec:ideal2D} that the solution
of (\ref{eq:2DHartreeFock}) can only accommodate a finite number of
atoms (\ref{eq:idealtrapped2}). However the situation is
dramatically changed in presence of repulsive interactions. Indeed
for any non zero $\gtwoD $, a solution to (\ref{eq:2DHartreeFock})
exists for arbitrarily large atom numbers \cite{Bhaduri:2000}.
Consequently no critical point can be found by simply searching for
a maximal atom number compatible with (\ref{eq:2DHartreeFock}). In
the following we will therefore turn to a different approach,
starting from the known exact (i.e. non mean-field) results
concerning the critical BKT point in a uniform interacting 2D Bose
 gas. The mean-field approximation will be used in a second
stage, in combination with the local density approximation (LDA), to
determine the critical atom number in the trapped system.

Note that it is also possible to pursue the search for a critical
point only within the mean-field approach, by looking for example
whether its solution exhibits a thermodynamical or dynamical
instability above a critical atom number
\cite{Fernandez:2002,Gies:2004a}. This instability would be an
indication that the system tends to evolve towards a different kind
of state, with a non-zero quasi-condensed and/or superfluid
component, and quasi-long range order
\cite{Petrov:2004a,Simula:2006}.

\subsection{The Berezinskii-Kosterlitz-Thouless transition
and the local density approximation}

In an  infinite uniform 2D Bose fluid, repulsive interactions have a
dramatic effect since they can induce a transition from the normal
to the superfluid state, when the temperature is lowered below a
critical value. The superfluid density jumps from 0 to $4/\lambda^2$
at the transition point \cite{Nelson:1977}. The microscopic
mechanism of the 2D superfluid transition has been elucidated by
Berezinskii \cite{Berezinskii:1971} and by Kosterlitz and Thouless
\cite{Kosterlitz:1973}. For a temperature larger than the critical
temperature, free vortices proliferate in the gas, destroying the
superfluidity. Below the transition, vortices exist only in the form
of bound pairs involving two vortices of opposite circulations,
which have little influence on the superfluid properties of the
system.

In a uniform system the phase space density $\DtwoD$ is a function
of the chemical potential and temperature $\DtwoD=F(\mu,T)$. For any
given $T$, the superfluid transition occurs when $\mu$ is equal to a
critical value $\mu_c(T)$. The corresponding critical value
$\DctwoD$ for the phase space density depends on the interaction
strength as \cite{Popov:1983,Kagan:1987b,Fisher:1988}
 \begin{equation}
\DctwoD=\ln (\xi/\tilde g)
 \label{eq:Prokofev}
 \end{equation}
 where $\xi$ is a dimensionless number. A recent Monte-Carlo
analysis provided the result $\xi=380\pm 3$ \cite{Prokofev:2001}
(see also \cite{AlKhawaja:2002}). For $\tilde g=0.13$ this gives a
critical phase space density $\DctwoD=8.0$.

\begin{figure}
\includegraphics[width=70mm]{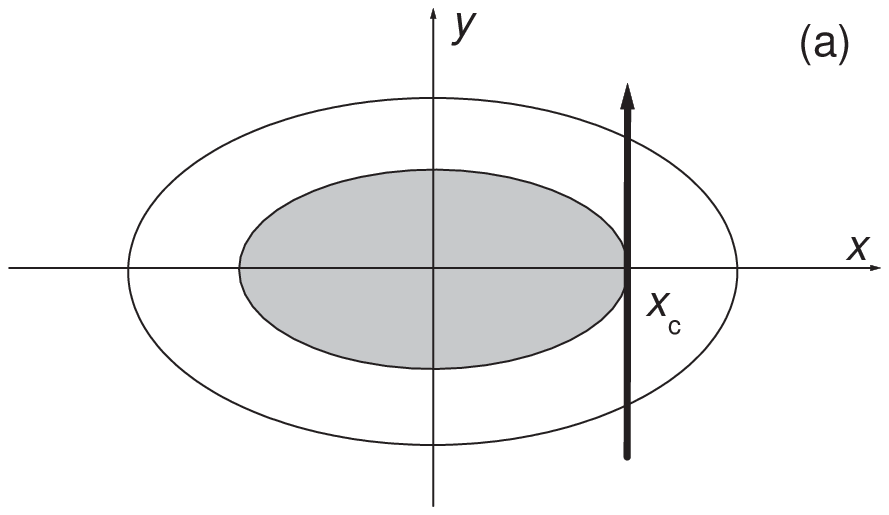}
\null \hskip 2cm
\includegraphics[width=70mm]{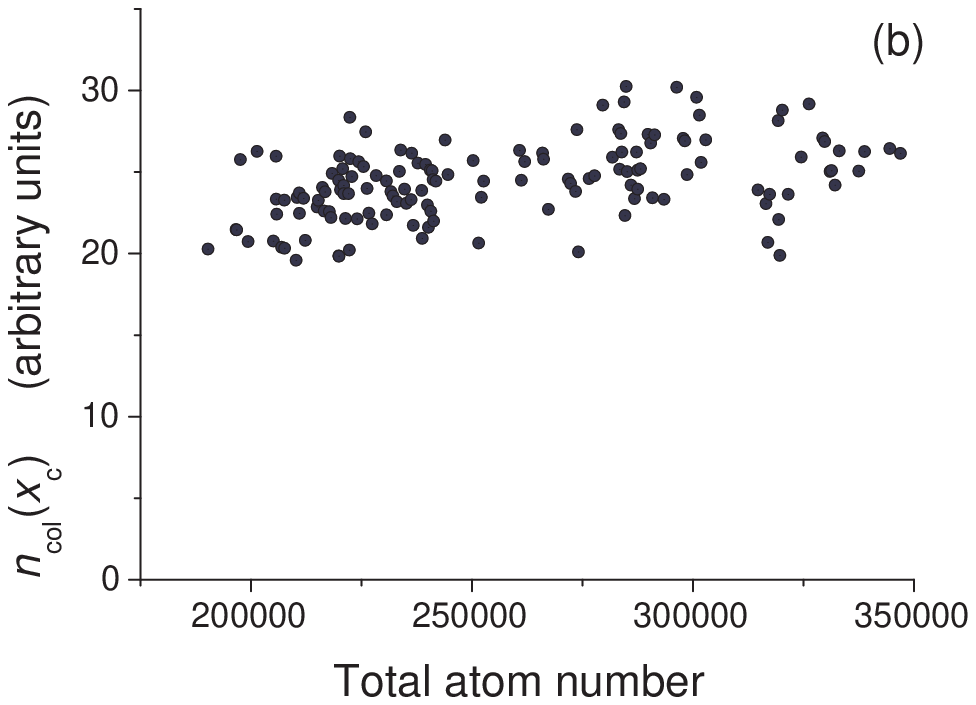}
\caption{Check of the local density approximation (LDA). (a) The
column density $n_{\rm col}$ is measured at the edge $x=x_{\rm c}$
of the central  part (in grey) of the bimodal distribution. (b)
$n_{\rm col}(x_{\rm c})$ is plotted as a function of the total atom
number $N$ in the harmonic trap + lattice configuration. Within the
LDA for a single plane, $n_{\rm col}(x_{\rm c})$ should be
independent of $N$, which is indeed nearly the case. The small
variation of $n_{\rm col}(x_{\rm c})$ for large $N$ may be due to
the appearance of a non negligible population in side planes of the
optical lattice potential. The data have been taken for the
effective temperature $T=105$~nK. Each point is extracted from a
single image.} \label{fig:LDAcheck}
\end{figure}

We consider now a trapped gas whose size is large enough to be well
described by the local density approximation (LDA). The phase space
density in ${\bf r}$ is given by $\DtwoD({\bf r})=F(\mu-V({\bf
r}),T)$ and a superfluid component forms around the center of the
trap if the central phase density $\DtwoD(0)$ is larger than
$\DctwoD$ \cite{Holzmann:2007a}.  The edge of the superfluid region
corresponds to the critical line where $\mu-V({\bf r})=\mu_{\rm c}$.
The phase space density along this line is equal to $\DctwoD$,
independently of the total number of atoms in the trap. This can be
checked experimentally and constitutes a validation of the LDA. The
integration of the experimental data along the line of sight $y$
does not lead to any complication because the trapping potential is
separable, $V({\bf r})=V_1(x)+V_2(y)$. Therefore the edges of the
superfluid region along the $x$ axis are located in $\pm x_{\rm c}$
such that $V_1(x_{\rm c})=\mu-\mu_{\rm c}$ (see figure
\ref{fig:LDAcheck}a), and the column density along the line of sight
passing in $x=x_{\rm c}$ is
 \begin{equation}
n_{\rm col}(x_{\rm c})=\frac{1}{\lambda^2}\int \DtwoD(x_{\rm
c},y)\;dy=\frac{1}{\lambda^2}\int F(\mu_{\rm c}-V_2(y),T)\;dy
 \end{equation}
which is also independent of the total atom number $N$. This is
confirmed experimentally, as shown in figure \ref{fig:LDAcheck}b
where we plot $n_{\rm col}(x_{\rm c})$ as a function of $N$.  The
slight increase (10\%) of $n_{\rm col}(x_{\rm c})$ for atom numbers
larger than $3\times 10^5$ may be due to the fact that the
population of additional planes becomes non-negligible for such
large $N$.

The possibility to use the LDA to study the
Berezinskii-Kosterlitz-Thouless critical point in a harmonically
trapped quasi-2D Bose gas has been checked recently by Holzmann and
Krauth using a quantum Monte-Carlo analysis \cite{Holzmann:2007b}.
For trap parameters close to the ones of \cite{Kruger:2007} they
have shown that a superfluid component, characterized by a reduced
moment of inertia, indeed appears at the center of the trap when the
local phase space density reaches a critical value  close to the
prediction (\ref{eq:Prokofev}).

\subsection{Density profile in the 2D mean-field theory}

\label{sec:trapped_interacting}

In this section we use the mean-field  Hartree-Fock approximation
(\ref{eq:2DHartreeFock}) to calculate the density profile of the
trapped atomic cloud. As we mentioned above, this equation admits a
solution for any value of the fugacity $Z$, and therefore for an
arbitrarily large number of particles. Rewriting
(\ref{eq:2DHartreeFock}) as
 \begin{equation}
\DtwoD({\bf r})=- \ln\left( 1-Ze^{-\tilde g \DtwoD({\bf
r})/\pi}e^{-\beta V({\bf r})} \right)
 \label{eq:HFforD}
 \end{equation}
we see that the value of $\DtwoD$ for any temperature and at any
point in space depends only on the parameter $R$ defined by
$R^2=(x/x_T)^2+(y/y_T)^2$, where $x_T=(\omega_x^2m\beta)^{-1/2}$ and
$y_T=(\omega_y^2m\beta)^{-1/2}$. The total atom number is given by
 \begin{equation}
N=\left(\frac{k_BT}{\hbar \bar \omega}\right)^2\;\int_0^\infty
\tilde \DtwoD(R)\;R\;dR
 \label{eq:atom_number}
 \end{equation}
where $\tilde \DtwoD(R)$ is the solution of the reduced equation
 \begin{equation}
\tilde \DtwoD(R)=- \ln\left( 1-Ze^{-\tilde g \tilde \DtwoD
(R)/\pi}e^{-R^2/2} \right)\ .
 \label{eq:intermediate}
 \end{equation}
Quite remarkably this result for $\tilde \DtwoD(R)$ neither depends
on the trap parameters, nor on the temperature. The only relevant
parameters are the fugacity $Z$ and the reduced interaction strength
$\tilde g$. The scaling of the atom number $N$ with the temperature
and the trap frequency in (\ref{eq:atom_number}) is therefore very
simple. In particular it does not depend on the trap anisotropy
$\omega_y/\omega_x$ but only on the geometric mean $\bar \omega$.

For atom numbers much larger than $\NctwoDid$ it is interesting to
note that the radial density profile deduced from the mean-field
equation (\ref{eq:HFforD}) exhibits a clear bi-modal shape, with
wings given by $\ntwo(r)\lambda^2 \simeq Z e^{-\beta V({\bf r})}$
and a central core with a Thomas-Fermi profile $2g\,\ntwo(r)\simeq
\mu-V({\bf r})$. However this prediction of a bi-modal distribution
using the Hartree-Fock approximation cannot be quantitatively
correct. Indeed the Hartree-Fock treatment assumes a mean-field
energy $2\gtwoD\ntwo$. The factor 2 in front of this energy
originates from the hypothesis that density fluctuations are those
of a gaussian field $\langle n^2 \rangle =2\;\langle n\rangle^2$.
Actually when the phase space density becomes significantly larger
than 1, density fluctuations are reduced and one approaches a
situation closer to a quasi-condensate in which $\langle n^2 \rangle
\sim \langle n\rangle^2$ \cite{Prokofev:2001}. Taking into account
this reduction could be done for example using the equation of state
obtained from a classical field Monte-Carlo analysis in
\cite{Prokofev:2002} .

\begin{figure}
\centerline{\includegraphics[width=70mm]{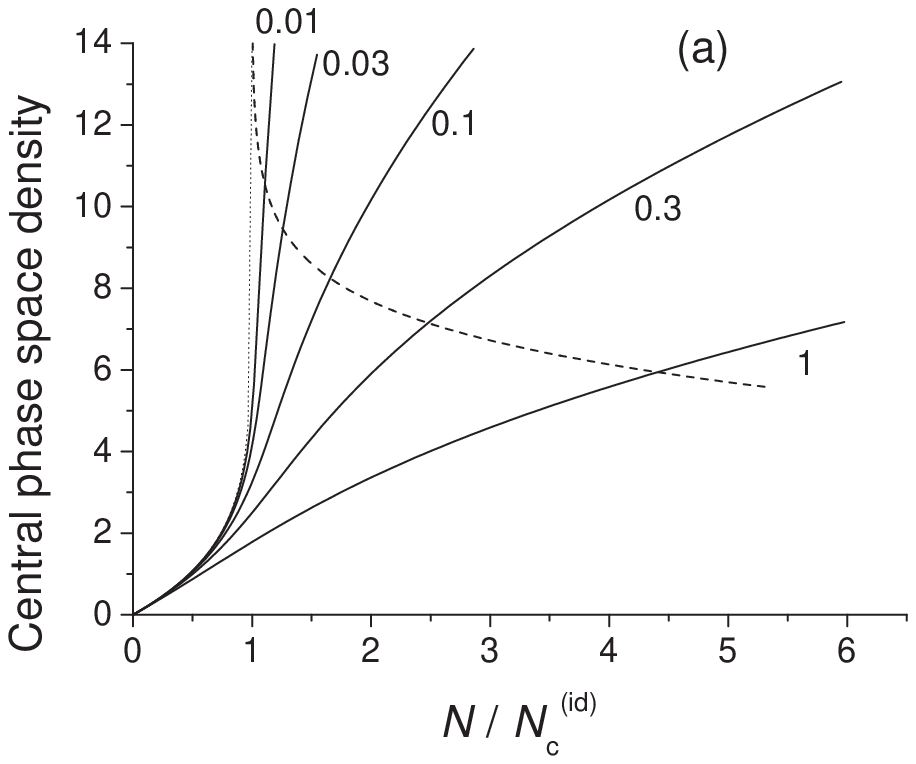}
\includegraphics[width=80mm]{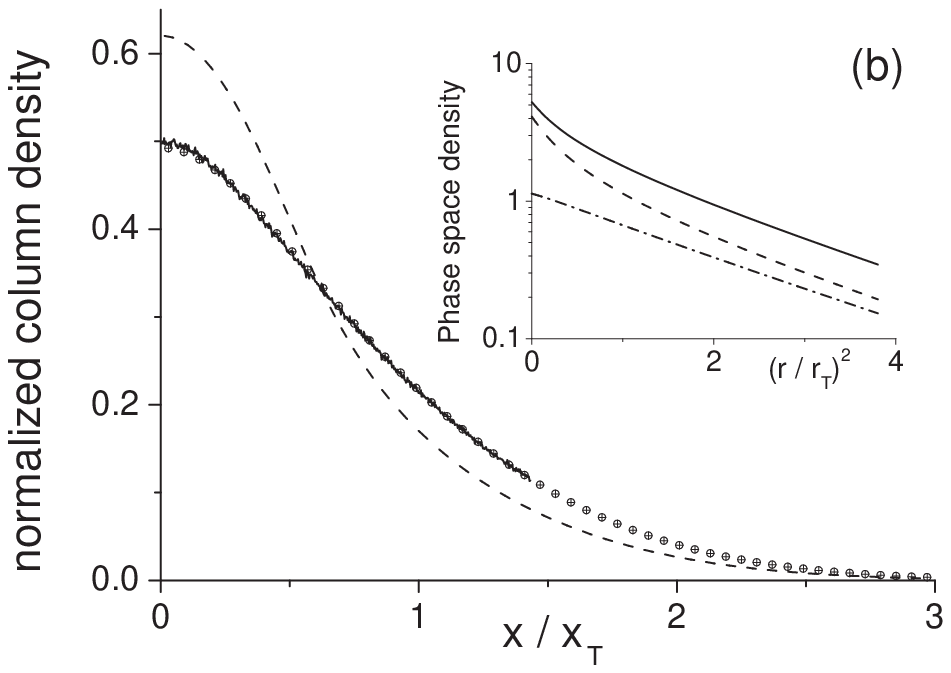}}
\caption{ (a) Central phase space density predicted by the 2D mean
field theory, as a function of the atom number for various
interaction strengths $\tilde g$. The dotted line represents the
semi-classical prediction for the ideal gas. The dashed line
indicates where  the threshold for superfluidity (\ref{eq:Prokofev})
is met at the center of the trap. (b) Column density for
$N=\NctwoDmf=36\,000$ atoms in an isotropic trap
($\omega_x=\omega_y=\omega$, $\tilde g=0.13$, $k_BT=110\,\omega$).
The dashed line is the result of the 2D mean-field analysis of
\S~\ref{sec:trapped_interacting}. The continuous line is the 3D
quantum Monte-Carlo result obtained in \cite{Holzmann:2007b}, with
$\omega_z=83\omega$ and $a=\tilde g a_z/\sqrt{8\pi}$. The dots are
the result of the hybrid 3D mean-field calculation of
\S~\ref{subsec:hybridMF} for the same parameters.  Inset: Prediction
of the hybrid 3D mean-field approach for the phase space density
(log scale) as a function of $r^2$ (a gaussian distribution leads to
a straight line). Continuous line: total phase space density; dashed
line: phase space density associated to the ground state $\varphi_1$
of the $z$ motion; dash-dot line: phase space density associated to
all other states $\varphi_j$, $j\geq 2$.} \label{fig:Nat_FSD}
\end{figure}

\subsection{Critical atom number in the 2D mean-field approach}
\label{subsec:critical2D}

We now use the solution of the mean-field equation (\ref{eq:HFforD})
to evaluate the critical atom number $\NctwoDmf$ that is needed to
reach the threshold (\ref{eq:Prokofev}) for the BKT transition at
the center of the trap $\DtwoD(0)=\ln(\xi/\tilde g)$. For a given
interaction strength $\tilde g$ we vary the fugacity $Z$ and  solve
numerically (\ref{eq:HFforD}) at any point in space. Examples of
spatial density profiles at the critical point are given in the
appendix, both before and after time-of-flight. The integration of
the density profile over the whole $xy$ plane gives the total atom
number $N$. From (\ref{eq:atom_number}) and (\ref{eq:intermediate}),
it is clear that the scaling of $\NctwoDmf$ with the frequencies
$\omega_{x,y}$ and with the temperature is identical to the one
expected for an ideal trapped gas.

We have plotted in figure \ref{fig:Nat_FSD}a the variation of
$\DtwoD(0)$ as a function of $N/\NctwoDid$ for various interaction
strengths. For a given atom number the phase space density at center
decreases when the strength of the interactions increases, as
expected. The numerical result for $\NctwoDmf$ is plotted as a
dashed line in figure \ref{fig:Nat_FSD}a. We find that it is in
excellent agreement -- to better than 1\%-- with the result of
\cite{Holzmann:2007a}
 \begin{equation}
\frac{\NctwoDmf}{\NctwoDid}=1+\frac{3\tilde g}{\pi^3} \DctwoD^2
 \label{eq:Ncanalytics}
 \end{equation}
over the whole range $\tilde g=0\,$--$\,1$.  This analytical result
was initially derived in \cite{Holzmann:2007a} for $g\ll 1$ using an
expansion around the solution for the ideal Bose gas, but this
approximation can actually be extended to an arbitrary value of
$\tilde g$ \cite{Holzmann:2007c}. The strongly interacting limit
($3\tilde g \DctwoD^2/\pi^3> 1$) can be easily understood by
noticing that in this case, the atomic distribution
(\ref{eq:2DHartreeFock}) nearly coincides with the Thomas-Fermi
profile $2gn(r)=\mu-V(r)$. Using the relation between the total atom
number and the central density for this Thomas-Fermi distribution
$N=2\pi \tilde g \left(\ntwo(0)\aoh^2\right)^2$ (with
$\aoh=(\hbar/(m\omega))^{1/2}$), one then recovers the second term
of the right-hand side of (\ref{eq:Ncanalytics}).

Let us emphasize that figure \ref{fig:Nat_FSD}a is a mix of two
approaches: (i) The mean-field model, that does not lead in itself
to  a singularity along the dashed line of figure
\ref{fig:Nat_FSD}a. (ii) The BKT theory for a uniform system, which
is a beyond mean-field treatment and which has been adapted to the
trapped case using the local density approximation in order to
obtain the critical number indicated by  the dashed line.

We now compare the 2D MF prediction with the results of the QMC
calculation of \cite{Holzmann:2007b}, looking first at the critical
atom number and then at the density profiles. For $\tilde g=0.13$
the mean field prediction for the critical number
(\ref{eq:Ncanalytics}) is $\NctwoDmf/\NctwoDid=1.8$. This is in
relatively good agreement with the QMC calculation of
\cite{Holzmann:2007b}, which gives $T_{\rm c}^{\rm (QMC)}=0.70\,
\TctwoDid$ or equivalently $\NctwoDQMC/\NctwoDid=2.0$. The QMC
calculation has been performed for various atom numbers $N$, for a
3D harmonic trap such that $\omega_z/\omega=0.43\sqrt{N}$ and a 3D
scattering length $a=\tilde g a_z/ \sqrt{8\pi}$, with $\tilde
g=0.13$.

The agreement between the 2D MF and the QMC approaches is not as
good for the density profiles close to the critical point. An
example is shown in figure \ref{fig:Nat_FSD}b, where we take
$k_BT=110\,\hbar \omega$ ($\NctwoDid=20\,000$). We choose
$N=\NctwoDmf=36\,000$ and we plot the column density $n_{\rm
col}(x)$ obtained by integrating the spatial density along the
directions $y$ and $z$. The mean field result is shown as a dashed
line, and it notably differs from the QMC result, plotted as a
continuous line. As we show below this disagreement is essentially a
consequence of the residual excitation of the $z$ motion
($k_BT/(\hbar \omega_z)=1.4$), that is neglected in the 2D MF
approach, whereas it is implicitly taken into account in the 3D QMC
calculation.

\subsection{The hybrid 3D mean-field approach}

\label{subsec:hybridMF}

In this section we extend the 2D mean-field treatment to take into
account the residual excitation of the $z$ motion. As pointed out to
us by the authors of \cite{Holzmann:2007b}, this is necessary for a
quantitative analysis of the experiment \cite{Kruger:2007}, since
the temperature and the chemical potential at the critical point
were not very small compared to $\hbar \omega_z^{\rm (lat)}$, but
rather comparable to it.

We follow here a method related to the one developed in section
\ref{sec:ideallattice} to analyze the ideal gas case. We start from
a trial 3D density distribution $\nthree({\bf r})$. At any point
$(x,y)$, we treat quantum mechanically the $z$ motion and solve the
eigenvalue problem for the $z$ variable
 \begin{equation}
\left[\frac{-\hbar^2}{2m}\frac{d^2}{dz^2}+ V_{\rm eff}({\bf r})
\right]\varphi_j(z|x,y)= E_j(x,y)\;\varphi_j(z|x,y)\ ,
 \end{equation}
where $V_{\rm eff}({\bf r})= m ({\omega}_x^2 x^2+\omega_y^2
y^2+\omega_z^2 z^2)/2\;+\;2 \gthreeD n_3({\bf r})$ and $ \int
|\varphi_j(z|x,y)|^2\;dz=1$. Treating semi-classically the $xy$
degrees of freedom, we obtain a new spatial density
  \begin{equation}
\nthree'({\bf r})=-\frac{1}{\lambda^2}\sum_j |\varphi_j(z|x,y)|^2
\ln \left(1-e^{\beta(\mu-E_j(x,y))} \right)\ .
  \end{equation}
We then iterate this calculation until the spatial density
$\nthree({\bf r})$ reaches a fixed point \cite{Kadanoff:1962}. With
this method, we fulfill two goals: (i) We take into account the
residual thermal excitation of the levels in the $z$ direction. (ii)
Even at zero temperature we take into account the deformation of the
$z$ ground state due to interactions.

This `hybrid 3D mean-field' method is different from the standard
mean-field treatment used to describe 3D Bose gases. In the latter
case, all three degrees of freedom are treated semi-classically,
which is valid when the particle population is distributed smoothly
over several quantum states. This standard 3D mean-field would not
be applicable in our case, where a significant part of the total
population accumulates in the lowest state $\varphi_1$.

An example of a result is shown in figure \ref{fig:Nat_FSD}b, where
we plot the column density $n_{\rm col}(x)$ obtained with this
hybrid MF method, taking into account the first 5 eigenstates
$\varphi_j$. The agreement between the hybrid 3D mean-field
prediction and the `exact' QMC prediction of \cite{Holzmann:2007b}
is excellent. This shows that the predictions of this hybrid 3D
mean-field approach are reliable as long as the superfluid
transition has not been reached at the center of the trap.

We show in the inset the variations of the phase space density
$D(r)$. We plot $\ln(D)$ as a function of $r^2$, so that a gaussian
distribution would appear as a straight line. The dashed line is the
phase-space density associated to the ground state of the $z$-motion
$\varphi_1$, and the dash-dot line corresponds to the excited states
$\varphi_j$, $j\geq 2$. The continuous line is the total phase-space
density. At the center of the trap, as a consequence of Bose
statistics, most of the population (80\%) accumulate in the ground
state $\varphi_1$. For $r \geq r_T$, the repartition of the
population among the eigenstates of the $z$ motion follows the
Boltzmann law, and $\sim 50\%$ of the atoms occupy the excited
states $\varphi_j$, $j\geq 2$. A practical consequence of this
increasing influence of excited states of the $z$-motion with
increasing $r$ is that the total phase space density profile is
notably closer to a gaussian distribution than when only the ground
state of the $z$-motion is retained in the calculation, as it is the
case in the 2D mean-field approach.

Finally we mention that we have also developed a simpler version of
this hybrid 3D mean-field analysis, in which the $\varphi_j$ levels
are not calculated self-consistently, but are assumed to coincide
with the energy eigenstates in the potential $m\omega_z^2 z^2/2$
(see also \cite{Holzmann:2008}). For the domain of parameters
relevant for the experiment, the two approaches lead to very similar
results.

\subsection{Mean-field approach for the lattice configuration and
comparison with experiment}

\label{subsec:latticeMF}

In order to compare the predictions of the mean-field approach with
the experimental results, we now turn to the lattice geometry,
corresponding to a stack of parallel planes located in
$z_j=z_0+j\ell$, $j$ integer. For simplicity we assume that the
laser waist $W_x$ is large compared to the spatial extent of the
atomic cloud, so that we can treat the gas as a superposition of
independent harmonically trapped systems. Each system is located at
the vicinity of a nodal plane of the optical lattice, and is treated
as a harmonic trap with frequencies
$\omega_x',\omega_y,\omega_z^{\rm (lat)}$. Note that we include here
`by hand' the renormalization $\omega_x\rightarrow \omega_x'$ of the
$x$ frequency due to the finiteness of the laser waist, that we
discussed in \S~\ref{subsec:renorm}. The magnetic trap adds an extra
confinement along the $z$ axis with a frequency $\omega_z$ so that
the chemical potential for the plane $z_j$ is $\mu_j=\mu-m\omega_z^2
(z_0+j\ell)^2/2$. We assume that the critical point is reached when
the phase space density associated to the lowest eigenstate
$\varphi_1$ in the most populated plane reaches the critical value
(\ref{eq:Prokofev}). Once the corresponding fugacity is determined,
we calculate the spatial distribution in each plane, sum up the
various contributions, and integrate the spatial distribution over
the line of sight $y$ to obtain the column density $n_{\rm col}(x)$.

\begin{figure}[t]
\centerline{\includegraphics[width=70mm]{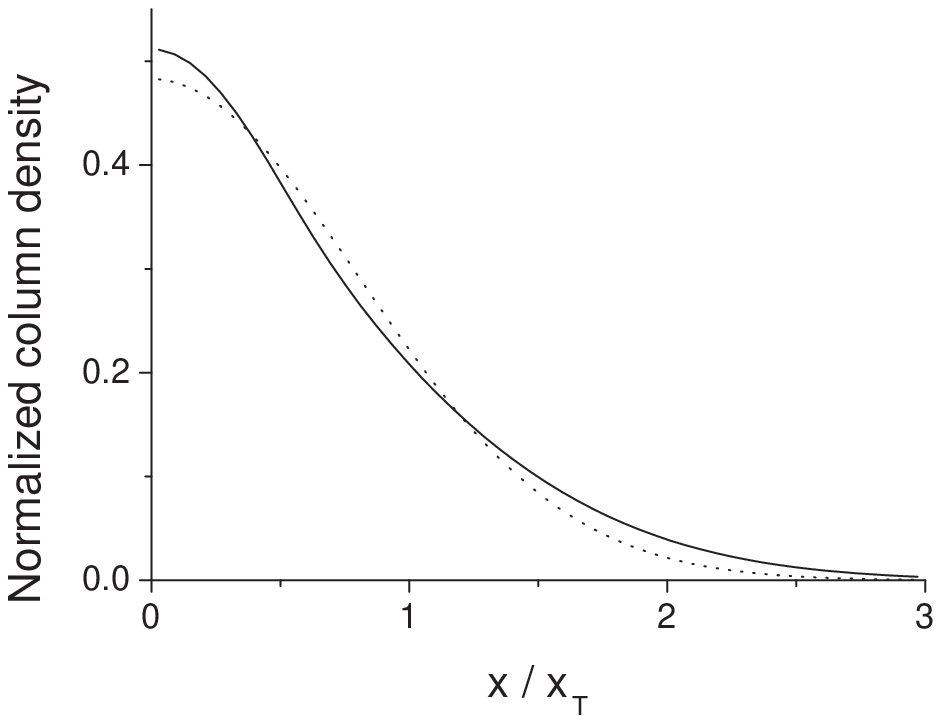}
\includegraphics[width=80mm]{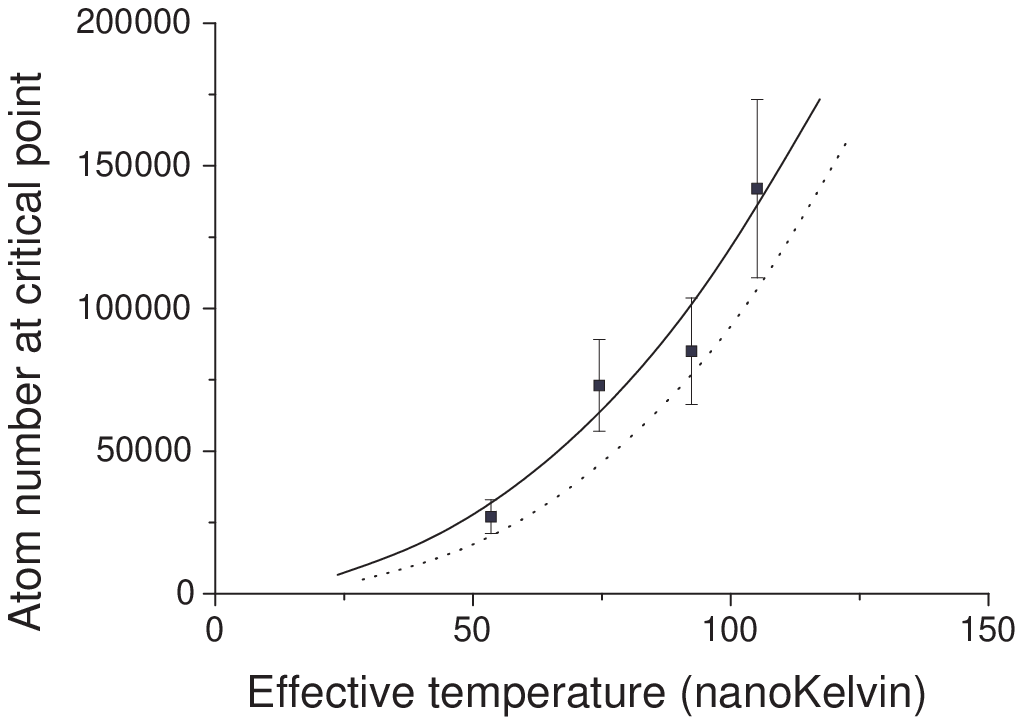}}
 \caption{(a) Hybrid 3D mean-field prediction for the
normalized column density for the lattice configuration A,
$T=150~nK$ and $N=110\,000$ atoms. For these parameters the phase
space density associated to the lowest eigenvalue of the $z$ motion
reaches the critical value $\DctwoD$ at the center of the most
populated planes. The dotted line is a gaussian fit which gives the
effective temperature $T_{\rm eff}=0.64\,T=96$~nK. (b) Critical atom
number as a function of the effective temperature obtained from a
gaussian fit of the MF result. The continuous line (resp. dashed)
line is for configuration A (resp. B). The points are the
experimental results of \cite{Kruger:2007}, already shown in fig.
\ref{fig:Nc_lattice}. } \label{fig:MFlattice}
\end{figure}

A typical result for $n_{\rm col}$ is given in figure
\ref{fig:MFlattice}a for the temperature $T=150$~nK and for the
lattice configuration A. The total atom number is $1.1\times 10^5$.
It is well fitted by a gaussian distribution $\exp(-x^2/2x_0^2)$
(dotted line), so that we can assign an effective temperature to
this distribution $T_{\rm eff}=m\omega_x^2 x_0^2/k_B$. In the
example of fig.\ref{fig:MFlattice}a, we find $T_{\rm eff}=0.64\,T$
(96~nK). For the same $T$ and a lattice in configuration B, the
effective temperature obtained with a gaussian fit is $T_{\rm
eff}=0.69\,T$ (103~nK) and the total atom number is $N=1.0\times
10^5$. We have repeated this procedure for temperatures $T$ in the
range 100-200~nK and consistently found the ratio $T_{\rm eff}/T$ in
the range $0.6$--$0.7$, with a quality of the gaussian fit similar
to what is shown in figure \ref{fig:MFlattice}a.

We have plotted in figure \ref{fig:MFlattice}b the calculated total
number of atoms in the lattice at the critical point, as a function
of the effective temperature deduced from the gaussian fit to the
column density. We have also plotted the experimental points of
\cite{Kruger:2007} already shown in figure \ref{fig:Nc_lattice}. We
remind the reader that the experimental `effective' temperature is
also deduced from a gaussian fit to the measured column density. One
reaches in this way a good agreement between the experimental
results and the hybrid 3D mean-field prediction. The predicted
density profiles with the 3D MF approach therefore provide a
satisfactory means for temperature calibration. They indicate in
particular that for the experiment \cite{Kruger:2007}, the effective
temperatures are typically $30$--$40\%$ below the real ones. To
improve on the comparison between theory and experiment, a more
controlled setup will be needed with an accurate independent
measurement of temperature, as well as the possibility of addressing
only a single or a fixed number of planes.

\section{Summary and concluding remarks}

In this paper we have analyzed the critical point of a trapped
quasi-2D Bose gas. We have shown that the experimental results of
\cite{Kruger:2007} are not in agreement with the ideal Bose gas
theory. The differences are found first at the qualitative level:
the predicted shape for the ideal gas distribution is `peaky' around
its center, which clearly differs from the quasi-gaussian measured
profile. Also the measured critical atom numbers $N_{\rm c}(T)$ do
not agree with the predictions for the ideal gas. Using the
`effective' temperatures obtained by treating the gaussian profiles
as Boltzmann distributions, the measured $N_{\rm c}(T)$ are larger
by a factor $\sim 5$ than the predicted ones. We then discussed the
predictions of a hybrid approach based on the local density
approximation. It combines the density profile calculated using a
mean-field Hartree-Fock treatment, and the known result for the
critical phase space density for the BKT transition in an infinite,
uniform 2D Bose gas \cite{Prokofev:2001}. We compared the
predictions of this approach with the results of a recent Quantum
Monte-Carlo calculation \cite{Holzmann:2007b} and reached the
following conclusions: (i) If one is interested only in the critical
atom number, it is sufficient to use a strictly 2D mean-field
approach. It leads to the approximate analytical result
(\ref{eq:Ncanalytics}), in good agreement with the QMC prediction.
For the experimental parameters of \cite{Kruger:2007} the critical
atom number is $N_c \sim 2\,\NctwoDid$.  (ii) In order to calculate
accurately the density profiles for the experimental temperature
range ($k_B T$ between $0.5\,\hbar \omega_z$ and $\hbar \omega_z$),
it is important to take into account the residual excitation of the
$z$ degree of freedom (the same conclusion has been reached in
\cite{Holzmann:2008}). We have presented a hybrid 3D mean-field
approximation which leads to density distributions in excellent
agreement with the QMC predictions close to the critical point. When
generalized to the lattice geometry used in the experiment, the
predicted density profiles are close to a gaussian distribution, and
a good agreement between theory and experiment is reached concerning
the critical number $N_c(T)$ when the predicted density profile is
used for temperature calibration.

\begin{figure}
\centerline{\includegraphics[width=80mm]{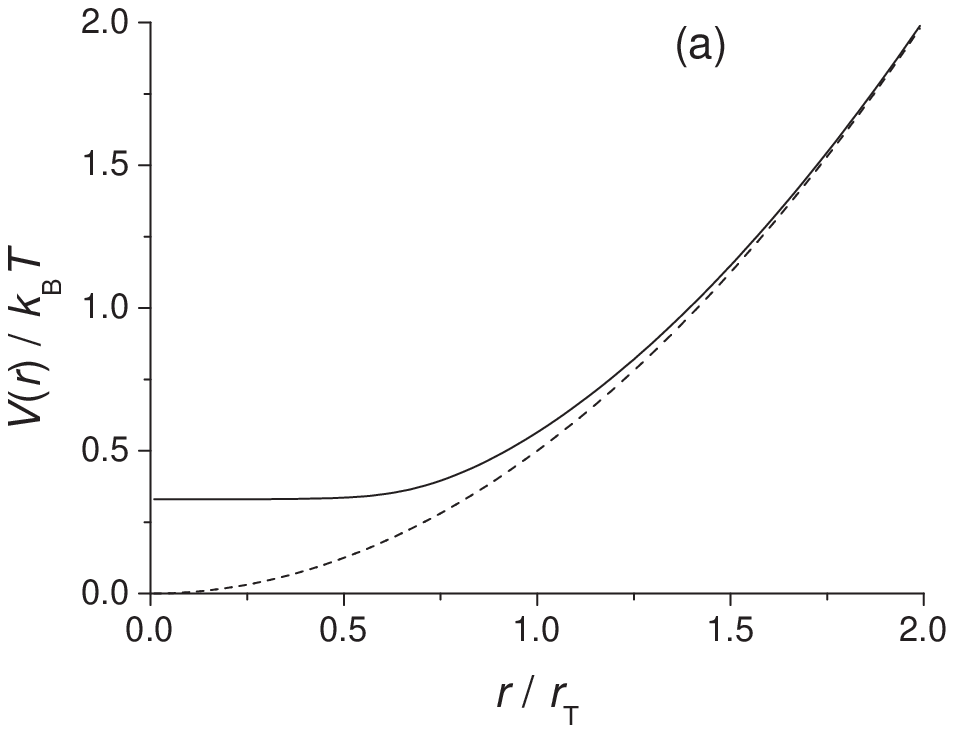}
\includegraphics[width=80mm]{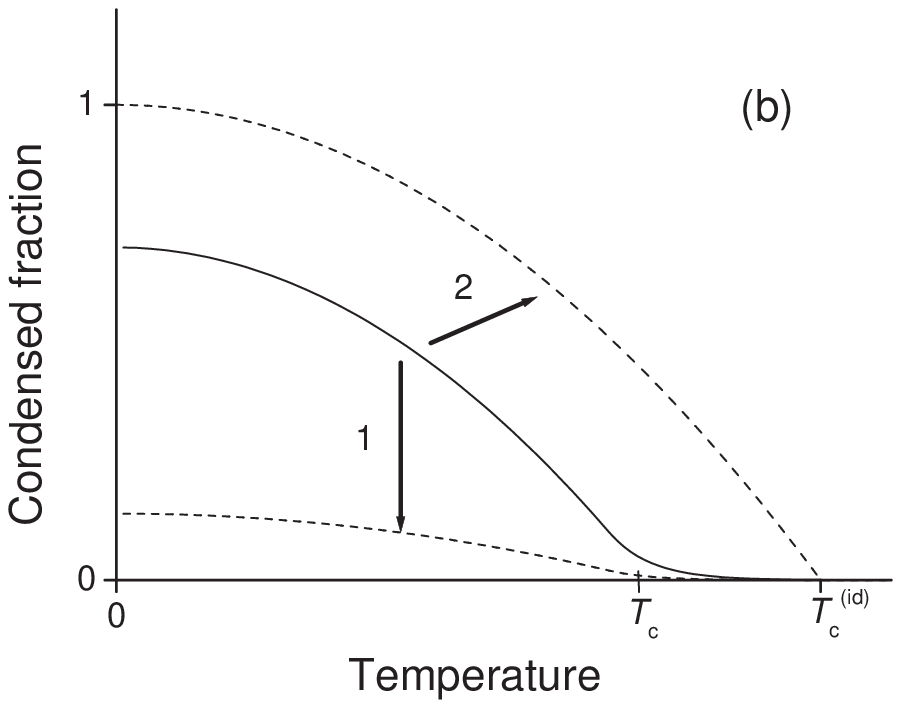}
} \caption{ (a) Trapping potential $V(r)$ in dashed line and
effective mean-field potential $V_{\rm eff}(r)=V(r)+2\gtwoD\ntwo(r)$
in continuous line, for $\tilde g=0.13$ and a central phase space
density equal to the critical value (\ref{eq:Prokofev}). (b)
Schematic representation of the condensed fraction in a finite 2D
Bose gas for a given interaction strength $\tilde g$ (continuous
line). Two limits can be considered: (1) thermodynamic limit $N\to
\infty$, $\omega\to 0$, $N\omega^2$ constant; the condensed fraction
tends to zero for any non zero value of $\tilde g$. (2) ideal gas
limit $\tilde g\to 0$.
 }
\label{fig:conclusion}
\end{figure}

We now briefly discuss the nature of the critical point that appears
in the trapped 2D Bose gas and compare it with `standard'
Bose-Einstein condensation. For a harmonically trapped ideal gas, we
recall that conventional Bose-Einstein condensation is expected in
the thermodynamic limit $N\to \infty$, $\omega\to 0$, $N\omega^2$
constant. This is a consequence of the density of states for a
quadratic hamiltonian around the zero energy. The price to pay for
this condensation in a 2D system is a diverging atomic density at
the center of the trap.  On the contrary when interactions are taken
into account, the mean-field approximation leads to the potential
$V({\bf r})+2\gtwoD\ntwo({\bf r})$ that is flat at the origin
(figure~\ref{fig:conclusion}a and \cite{Holzmann:2007a}). The
`benefit' of the harmonic trapping potential is lost and the physics
of the trapped interacting gas is very similar to that of a uniform
system. In particular one expects in the thermodynamic limit the
appearance of quasi-long range order only, with no true
Bose-Einstein condensate \cite{Petrov:2000a}. \footnote{A similar
flattening of the mean-field potential occurs in 3D, but it has no
important consequence in this case since true BEC is possible in an
infinite, uniform 3D system.} The transition between the ideal and
the interacting case is explicit in equations (\ref{eq:Prokofev})
and (\ref{eq:Ncanalytics}), where the limit $\tilde g\to 0$ gives
$\DctwoD\to +\infty$ and $\NctwoDmf/\NctwoDid \to 1$. In particular
(\ref{eq:Ncanalytics}) can be used to separate a `BEC-dominated'
regime where $\eta=3\tilde g \DctwoD^2/\pi^3 \ll 1$ and $N_{\rm
c}\simeq \NctwoDid$, and a `BKT-dominated' regime, where the
contribution of $\eta$ is dominant and $N_{\rm c}\gg \NctwoDid$. In
the latter case, the spatial distribution in the mean-field
approximation is a Thomas-Fermi disk with radius $R_{\rm TF}$ and
(\ref{eq:Ncanalytics}) is equivalent (within a numerical factor) to
the BKT threshold (\ref{eq:Prokofev}) for a uniform gas with density
$n=N/(\pi R_{\rm TF}^2)$. The rubidium gas studied in
\cite{Hadzibabic:2006,Kruger:2007} is at the border of the
`BKT-dominated' regime ($\eta \simeq 1$), whereas previous
experiments performed on quasi-2D gases of sodium atoms
\cite{Gorlitz:2001b} corresponded to $\eta\sim 0.1$, well inside the
`BEC-dominated' regime.

Finally, we must take into account the finite size of the gas in our
discussion. It is known from simulations of 2D spin assemblies that
for a finite size system, the average magnetization increases
rapidly around the BKT transition \cite{Bramwell:1994}. It is at
first sight surprising that this magnetization can be used as a
signature of BKT physics, since it would not exist in an infinite
system where a genuine BKT transition takes place. However it is
relevant for all practical 2D situations: as emphasized in
\cite{Bramwell:1994} one would need extremely large systems (`bigger
than the state of Texas') to avoid a significant magnetization even
just below the transition point. A similar phenomenon occurs for a
finite size Bose gas. A few states acquire a large population around
the transition point, and this allows for the observation of good
contrast interferences between two independent gases. In particular
the condensed fraction $f_0$ (largest eigenvalue of the one-body
density matrix) is expected to grow rapidly at the critical point,
and this has been quantitatively confirmed by the QMC calculation of
\cite{Holzmann:2007b}. To illustrate this point we have
schematically plotted in figure~\ref{fig:conclusion}b the expected
variations of $f_0$ with the parameters of the problem. For given
$\tilde g$ and $N$, $f_0$ takes significant values for $T<T_{\rm c}$
(continuous line). If the strength of the interactions $\tilde g$ is
kept constant, the condensed fraction $f_0$ tends to zero for any
finite temperature if the thermodynamic limit is taken (arrow 1 in
figure~\ref{fig:conclusion}b). Note that the superfluid fraction
should tend to a finite value in this limiting procedure. Now one
can also keep $N$ constant and decrease $\tilde g$ to zero (arrow 2
in figure~\ref{fig:conclusion}b). In this case one expects to
recover the ideal gas result $f_0=1-(T/\TctwoDid)^2$ for any value
of $N$. Therefore we are facing here a situation where two limits do
not commute: $\lim_{N\to \infty}\lim_{g\to 0}\neq \lim_{g\to 0}
\lim_{N\to \infty}$. Of course this does not cause any problem in
practice since none of these limits is reached. In this sense the
phenomenon observed in our interacting, trapped 2D Bose gas is
hybrid: the transition point is due to BKT physics (the density of
states of the ideal 2D harmonic oscillator does not play a
significant role because of the flattening of the potential), but
thanks to the finite size of the system, some diagnoses of the
transition such as the appearance of interferences, take benefit of
the emergence of a significant condensed fraction.

\ack We are indebted to Markus Holzmann and Werner Krauth for
numerous discussions, for providing the Quantum Monte-Carlo data
shown in figure \ref{fig:Nat_FSD}, and for pointing out the
significant role of the excitation of the $z$ motion in our
experiments. We also thank Yvan Castin and Pierre Clad\'e for
helpful discussions. P.K. and S. P. R. acknowledge support from EU
(contract MEIF-CT-2006-025047) and from the German Academic Exchange
Service (DAAD, grant D/06/41156), respectively. This work is
supported by R\'egion \^Ile de France (IFRAF), CNRS, the French
Ministry of Research, ANR (Project Gascor, NT05-2-42103) and the EU
project SCALA. Laboratoire Kastler Brossel is a research unit of
\'Ecole Normale Sup\'{e}rieure, Universit\'e Pierre and Marie Curie
and CNRS.

\appendix

\section*{Appendix: The time-of-flight in the 2D mean-field approximation}

\setcounter{section}{1}

We have emphasized in this paper that the measured density profiles
differ from those calculated for an ideal gas or within  the 2D
mean-field theory. The profiles calculated in steady-state in the
trap are found much `peakier' than the experimental ones. As the
experimental profiles were actually measured after a time of flight
of $t=22$~ms ($\omega_x t=1.3$), it is important to check that this
mismatch between predicted and observed profiles remain valid when
the ballistic expansion of the atoms is taken into account. Also the
atom distributions were measured using an absorption imaging
technique, with an imaging beam propagating along the $y$ axis.
Therefore the measurement gave access to the column density $n_{\rm
col}(x,t)$, obtained by integrating the total density along $y$. In
this appendix we take into account the time-of-flight and the
integration along the $y$ direction, both for an ideal and for an
interacting gas within the 2D mean-field approximation.

\begin{figure}
\centerline{\includegraphics[width=80mm]{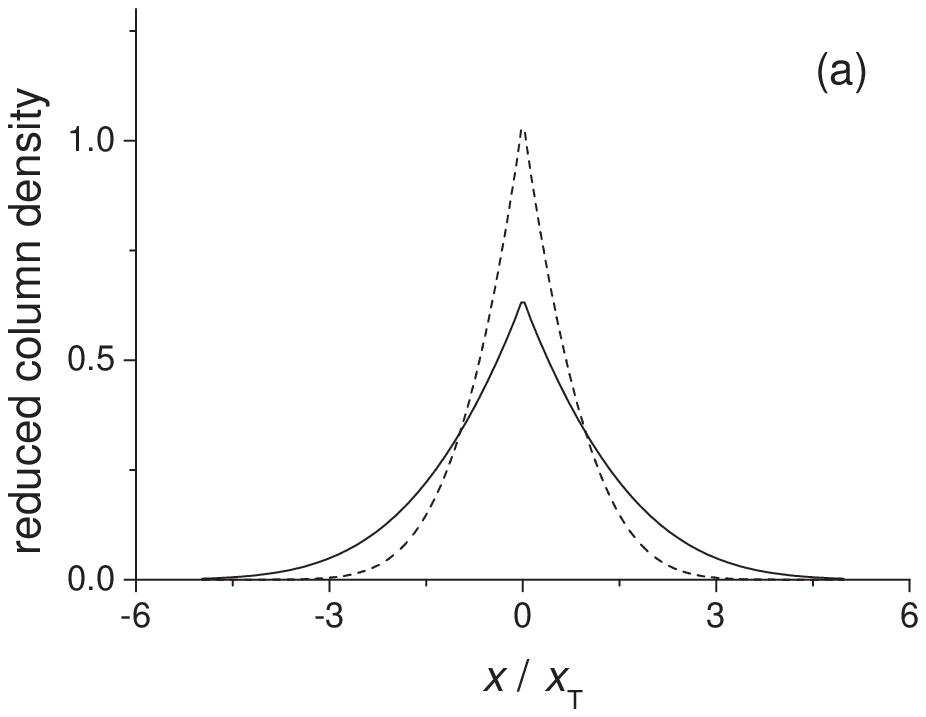}
\includegraphics[width=80mm]{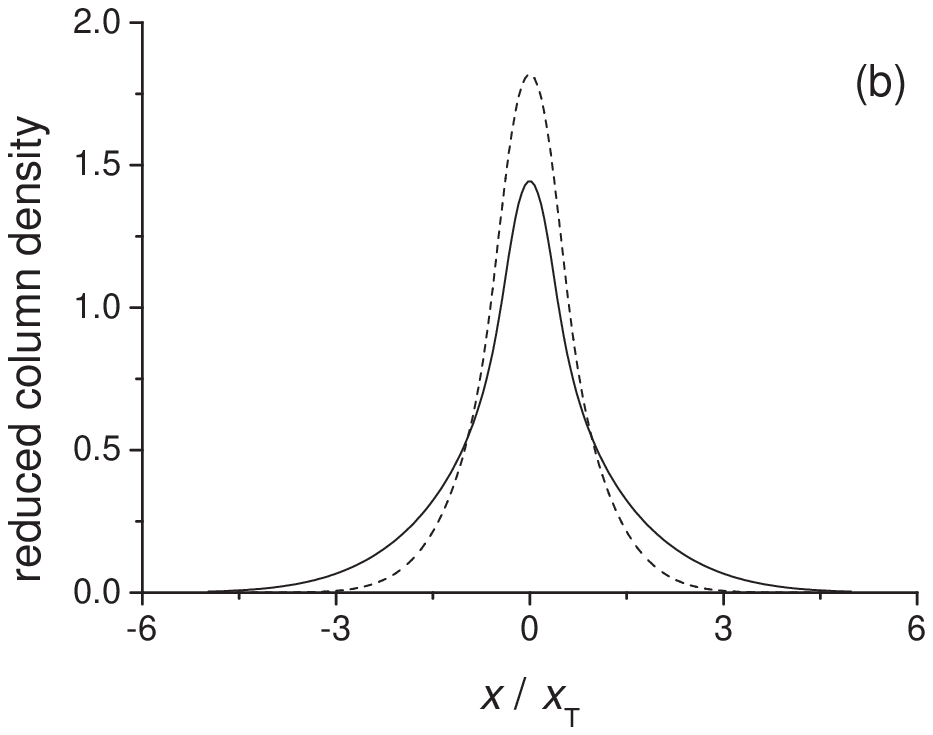}}
\caption{Reduced column density $F$ in the trap (dashed line) and
after a time of flight $t$ such that $\omega_xt=1.3$ (continuous
line). (a) Ideal gas case, for an atom number equal to the critical
value (\ref{eq:idealtrapped2}). (b) Mean-field result for $\tilde
g=0.13$. The fugacity is chosen such that the threshold for
superfluidity (\ref{eq:Prokofev}) is met at the center of the trap.
} \label{fig:TOF}
\end{figure}

The spatial distribution $\ntwo({\bf r},t)$ at time $t$ can be
calculated from the phase space distribution $\rhotwo({\bf r},{\bf p
})$ at initial time using
 \begin{equation}
\ntwo({\bf r},t)=\int \rhotwo({\bf r}-{\bf p }t/m,{\bf p })\; d^2p\
.
 \end{equation}
In the semi-classical approximation the in-trap phase space density
is given by
 \begin{equation}
\rhotwo({\bf r},{\bf p })=\frac{1}{h^2}\left\{
 \exp\left[ \left(p^2/(2m)+V_{\rm eff}({\bf r})-\mu
 \right) /k_BT\right]\; -\;1
\right\}^{-1}\ ,
 \end{equation}
where $V_{\rm eff}=V({\bf r})+2g\ntwo({\bf r})$, and $\ntwo({\bf
r})$ is obtained by solving (\ref{eq:HFforD}). The result for the
column density can be written as
 \begin{equation} n_{\rm
col}(x,t)=\frac{1}{x_T}\;\left( \frac{k_BT}{\hbar \omega}
\right)^2\;F(X,Z,\tilde g,\tau )\ , \quad X=\frac{x}{x_T}\ ,\quad
\tau=\omega_x t.
 \end{equation}
The results for $F$ are shown in figure \ref{fig:TOF}a for an ideal
gas, and in figure \ref{fig:TOF}b for an interacting gas in the mean
field approximation. In the ideal gas case, the initial column
density can be calculated analytically~:
 \begin{equation}
F(X,Z,0,0)=\frac{1}{\sqrt{2\pi}}\;g_{3/2}\left(Ze^{-X^2/2}\right)
 \end{equation}
and the column density after time of flight is deduced from the
initial value by a simple dilatation
 \begin{equation}
F(X,Z,0,\tau)=\frac{1}{\sqrt{1+\tau^2}}F\left(
\frac{X}{\sqrt{1+\tau^2}},Z,0,0\right)\ .
 \end{equation}
In figure \ref{fig:TOF}a, the fugacity is such that the atom number
equals the critical number (\ref{eq:idealtrapped2}). In the
interacting case of figure \ref{fig:TOF}b, the number of atoms is
such that the criterion for superfluidity is met at the center of
the trap. In all cases, it is clear that the observed profiles are
very different from a gaussian, in clear disagreement with the
experimental observation.

\vskip 1cm

\bibliographystyle{unsrt}

\end{document}